\documentclass{aa}  

\usepackage{graphicx}
\usepackage{txfonts}
\usepackage{hyperref}
\newcommand{\xmm}{{\em XMM-Newton}}

\newcommand{\kori}{$\kappa$~Ori}
\newcommand{\pn}{{\em pn}}

\newcommand{\lxu}{{erg~s$^{-1}$}}

\begin{document}

   \title{XMM-Newton imaging of V1818 Ori: a young stellar group on the eastern edge of the Kappa Ori ring.
          \thanks{Based on observations obtained with XMM-Newton, an ESA science mission with instruments 
          and contributions directly funded by ESA Member States and NASA} }

   \author{I. Pillitteri \inst{1} \and S. J. Wolk\inst{2} \and S. T. Megeath\inst{3} }

   \institute{INAF-Osservatorio Astronomico di Palermo, Piazza del Parlamento 1, 90134 Palermo, Italy\\
              \email{pilli@astropa.inaf.it}
         \and
             Harvard-Smithsonian Center for Astrophysics, 60 Garden St, Cambridge (MA), 02138 USA
         \and
             Ritter Astrophysical Research Center, Dept. of Physics and 
             Astronomy, University of Toledo, Toledo, OH, USA
             }

   \date{Received; accepted }

 
  \abstract
{We present the results of a 40 ks \xmm\ observation centered on the variable star V1818 Ori.
Using a combination of the \xmm\ and {\em All}WISE catalog data, we identify a group of about 
31 young stellar objects around V1818 Ori. This group is coincident with the eastern edge of the dust ring surrounding \kori. 
Previously, we concluded that the young stellar objects on the western side of ring were formed in an episode of 
star formation that started 3--5~Myr ago, and are at a distance similar to that of kappa Ori ($250-280$ pc) and 
in the foreground to the Orion A cloud.  
Here we use the \xmm\ observation to calculate X-ray fluxes and luminosities of the young stars around 
V1818 Ori. We find that their X-ray luminosity function (XLF), calculated for a distance of 
$\sim270$ pc, matches the XLF of the YSOs west of \kori. We rule out that this group of young stars is
associated to Mon R2 as assumed in the literature, but rather they are part of the same \kori's ring 
stellar population.
}


   \maketitle
%

\section{Introduction}
Understanding the spatial structure and distribution of young stellar objects (YSOs) in 
the Orion molecular cloud complex is crucial for tracing its star formation history.
The Orion OB association is characterized by several subgroups of stars of different masses, ages and distances;
{  its southern portion, Orion A, has distances between $\sim388$ pc and $428$ pc 
\citep{Kounkel2017, Bally08, Menten07}.} 
The Orion region thus offers a wide spectrum of test cases 
for various mechanisms of star formation, e.g., self collapse vs. triggered star formation, interplay 
of external agents as opposed to in situ evolution, or formation of low mass groups in isolation 
vs. mass segregated formation in denser and more massive clusters.
  
\begin{figure}
\centering 
\resizebox{0.49\textwidth}{!}{\includegraphics{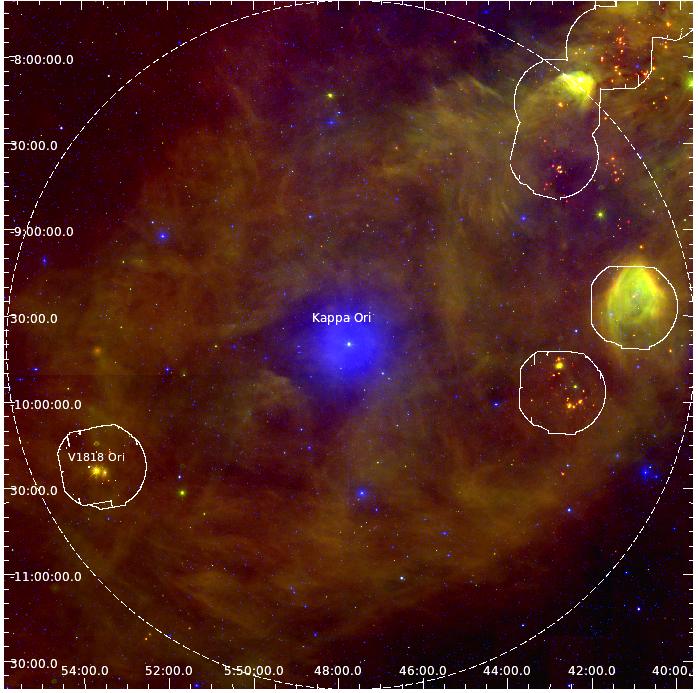}}
\caption{\label{wisergb}  
Composite RGB Optical (blue) + WISE 12$\mu$m (green) and 22$\mu$m (red) centered on \kori.
With white contours we mark on the right side the \xmm\ fields observed in \citet{Pillitteri2016} 
and \citet{Pillitteri2013}, and to the left the field around V1818 Ori. 
The ring of warm dust is visible in the WISE 
bands and extends $\sim2\deg$ from \kori\ (dashed large circle). 
}
\end{figure}

In this context, we investigated the X-ray emission of YSOs
west of \kori\ and to the South of Orion A/L1641 \citet{Pillitteri2016}. 
These young stars are within a ring of gas and dust apparently surrounding the B0 (V) star \kori\ 
within a radius of $2\deg$  (Fig. \ref{wisergb}).  
The ring appears as the rim of a bubble created by the sweeping action of the stellar winds emanating 
from \kori.
This structure overlaps the L1641 region of the Orion A cloud. 
However, the quoted distance to \kori\ is about $200-240\pm40$ pc \citep{Megier2009,VanLeeuwen2007}, 
much closer than Orion A (around 400 pc). 
We thus investigated the relationship between the YSOs in the ring and Orion A.
By means of X-ray observations of the YSOs in the ring 
of dust and the comparison of their X-ray luminosity function to that of L1641 and ONC stars, we determined 
that the distance to the YSOs near \kori\ is about $250-280$ pc, similar to the \kori\ distance, 
and as such they are unrelated to Orion A. 
They are rather a distinct population of YSOs born within the ring of gas and dust centered on \kori\ 
visible in extinction maps and mid IR and sub-mm images.

V1818 Ori is a Herbig Be star surrounded by about two dozen of YSOs and apparently sitting on 
the eastern edge of the same ring of dust centered on \kori\ (Fig. \ref{wisergb}).
V1818 Ori has highly irregular variability and bursts \citep{Chiang2015}, and
it is a binary system with components separated by $3.5\arcsec$. 
The spectra show strong veiling perhaps due to active accretion, and the stellar 
variability could also affect the determination of its spectral type, 
which is B7 as reported by \citet{Chiang2015}  but it is F0 as classified by \citet{Connelley2010}. 
Based on the similarity of velocity of the stellar absorption lines with that of the gas towards the Mon R2 association, 
 \citet{Chiang2015} assumed that these YSOs are part of the Mon R2 star forming complex at a distance of 900 pc. 
However, given its coincidence with the \kori' ring, the issue of the distance to V1818 Ori remains unsolved.

In the present paper we analyze an \xmm\ observation centered on V1818 Ori to resolve 
the issue of the distance to V1818 Ori and its nearby YSOs, 
to search for evidence of young stars to the east of \kori\ and to establish the extent 
and spatial structuring of the ongoing star formation in the ring.

The paper is structured as following: in Sect. \ref{observation} we present the observation and the data analysis,
in Sect.  \ref{results} we present our results and we discuss them, 
in Sect.  \ref{conclusions} we present our conclusions. 

\section{Observation and data analysis \label{observation}}
\begin{figure}
\centering 
\resizebox{0.49\textwidth}{!}{\includegraphics{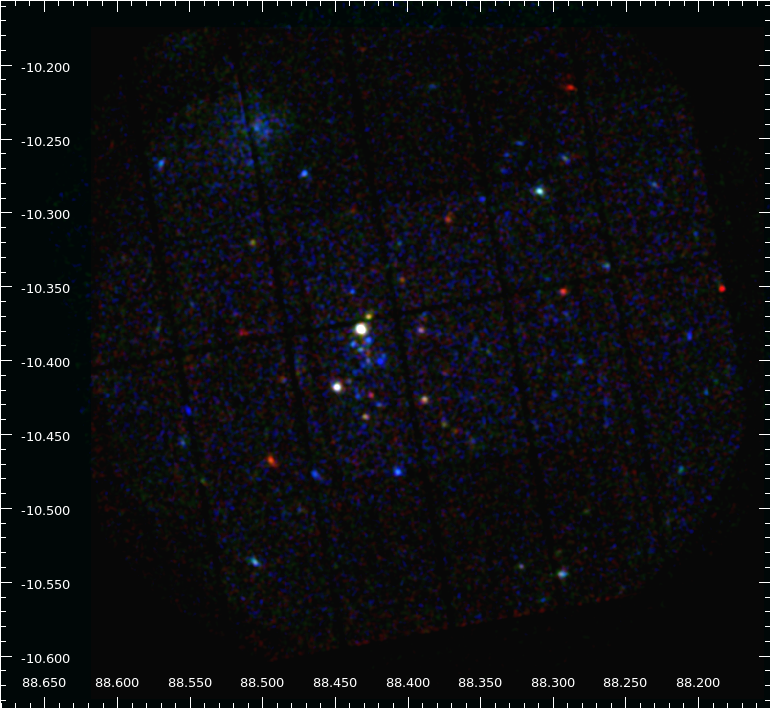}}
\caption{\label{rgb} Composite RGB image of V1818 Ori as observed with \xmm. The color bands are: 
0.3-1.0 keV (red), 1.0-2.5 keV (green), and 2.5-5.0 keV (blue). About 91 point like sources and two 
faint extended sources are detected in the image.}
\end{figure}

We observed {  the region towards V1818 Ori} with \xmm\ on April 6 2017 for a nominal duration of 40 ks 
(PI: Ignazio Pillitteri, ObsId 0800190101). 
The nominal pointing was at R.A.= $05^h53^m30^s$, Dec.=$-10^d22^m30.8^s$ (J2000) about 3 arcmin
offset from the optical position of V1818 Ori. We used EPIC as a prime instrument and the {\em Medium} filter. 
Figure \ref{rgb} shows a composite RGB image of the \xmm\ observation in three bands: 
0.3-1.0 keV (red), 1.0-3.0 keV (green), and 3.0-8.0 keV (blue).

We used SAS ver. 15 to reduce the data and obtain tables of events calibrated in astrometry, energy and timing. 
We selected the events of MOS 1, MOS 2 and \pn\ in the 0.3-8.0 keV band, with {\sc FLAG = 0} and {\sc PATTERN <=12} 
as prescribed by the SAS guide.

The background was highly variable during the observation and, in order to maximize the signal to noise of faint sources,
we used the light curve at high energies ($E>10$ keV) and a cut in rate to select good time intervals and filter out 
intervals with high background. At this step we removed about 15 ks leaving thus 25 ks of low background exposure, 
this is the portion of exposure time that we used for the subsequent source detection process. 
The source detection was made with a wavelet convolution technique implemented in a FORTRAN code that 
finds local maxima in the wavelet convolved image at different wavelet scales \citep{Damiani1997b, Damiani1997a}. 
We used a threshold of 4.6 $\sigma$ of local background to identify local maxima as X-ray sources, this value 
should retain, at most, one spurious source in the image due to statistical fluctuations of the background.
We detected 91 point-like sources; we can also recognize by eye a couple of faint extended sources to the north east 
corner of the image that have two radio galaxies as likely counterparts. 

\begin{figure}
\centering 
\resizebox{0.5\textwidth}{!}{
               \includegraphics{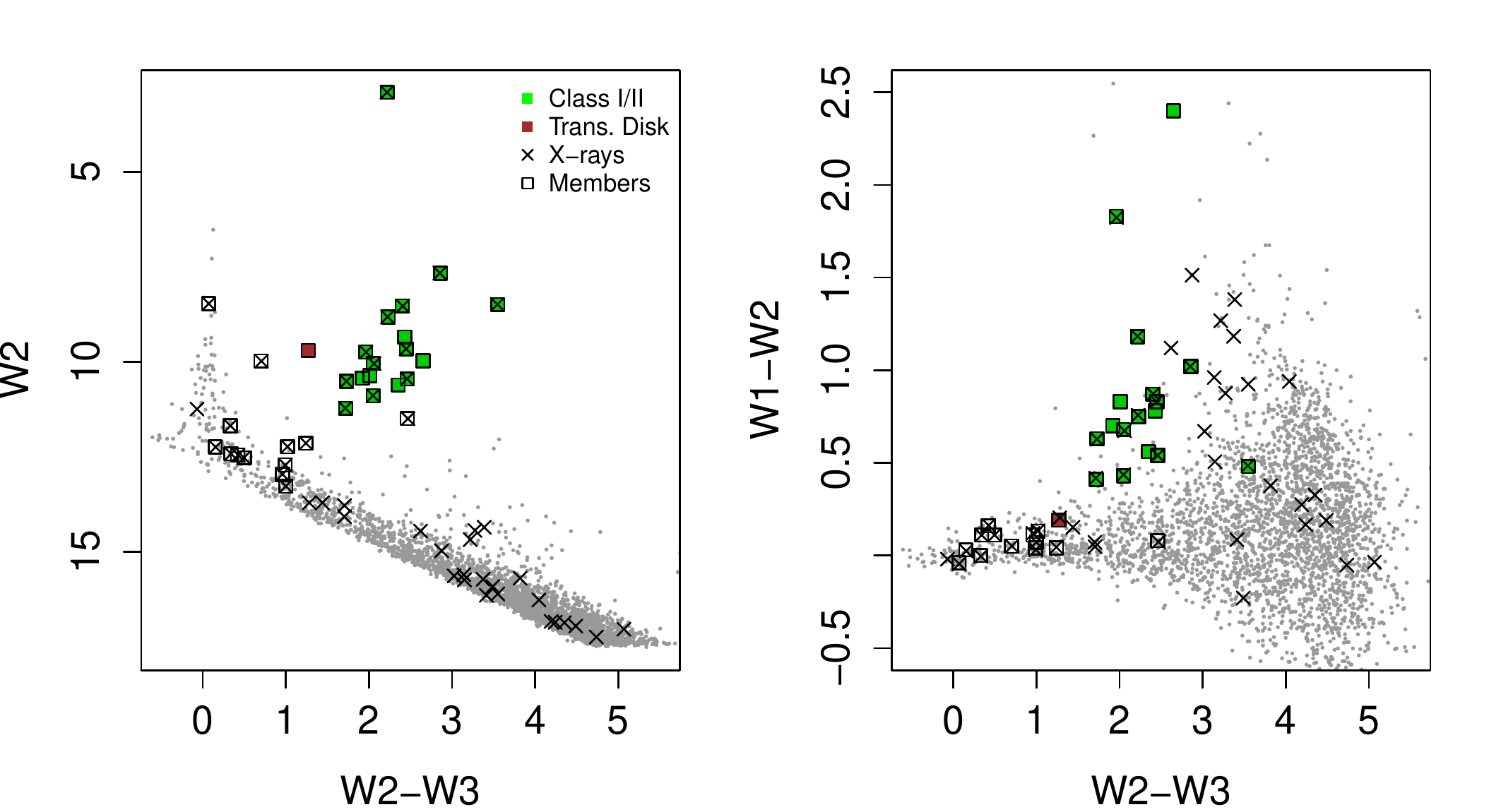}}
\resizebox{0.5\textwidth}{!}{
               \includegraphics{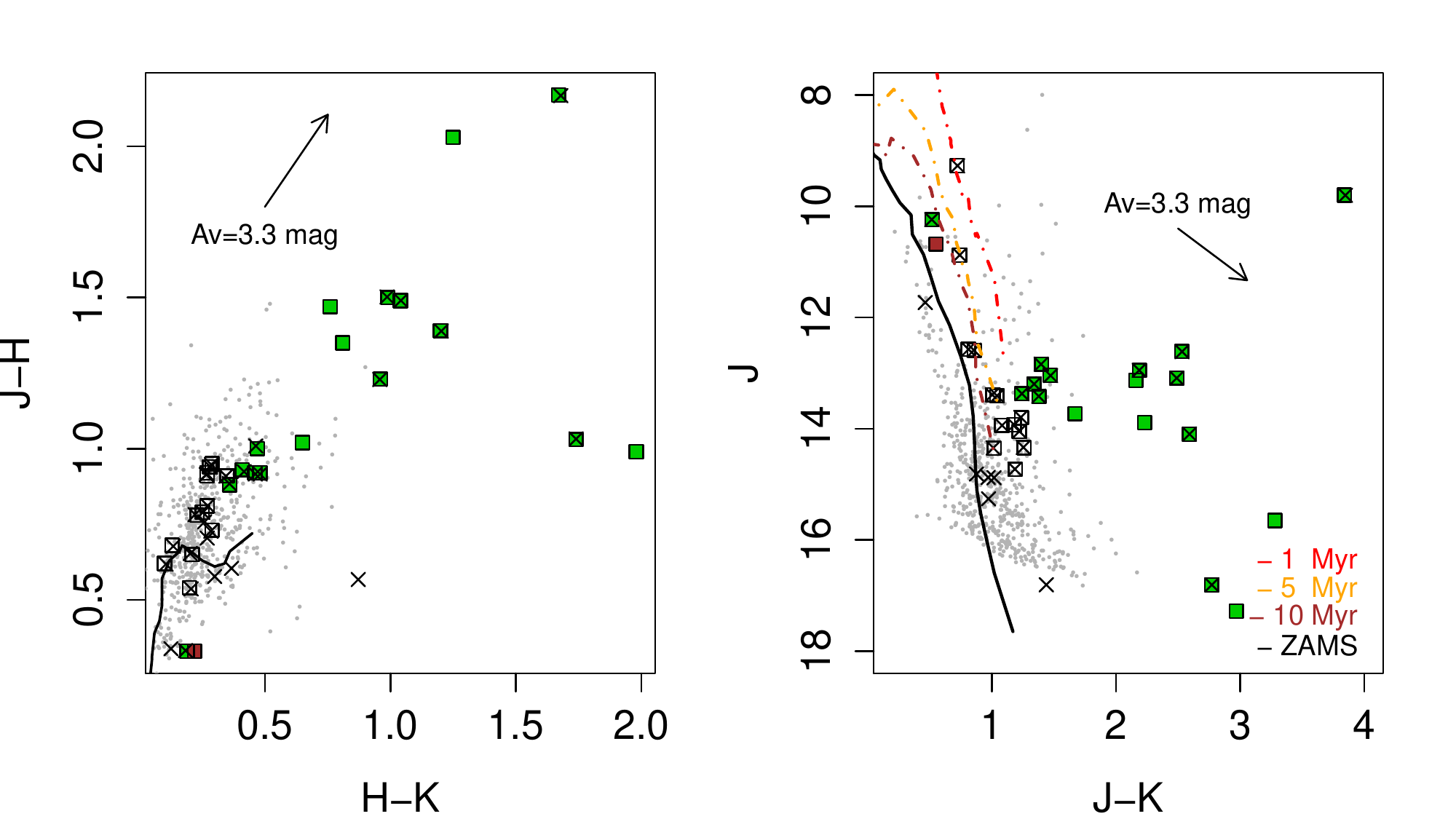}}
\caption{\label{ircmd} Top panels: WISE color-color and color-magnitude diagrams of the objects in the \xmm\
field of view. W1, W2 and W3 refer to the 3.4 $\mu$m, 4.6 $\mu$m and 12 $\mu$m WISE bands.
Selected members of the V1818 Ori group are open squares.
Bottom panels: 2MASS color-color and color-magnitude diagrams.  
Solid line represents the zero age Main Sequence, dashed lines are isochrones at 1, 5 and 10 Myr, respectively, 
for a distance of 270 pc (distance moduli for 270 pc, 400 pc and 900 pc are $\sim7.2$mag,$\sim8$ mag and $\sim9.1$ mag,
respectively). The reddening vector corresponding to A$_\mathrm V = 3.3$~mag represents the average extinction in the \xmm\ field.
}
\end{figure}

{ 
We searched for infrared counterparts to the X-ray sources in the {\em All}WISE catalog \citep{Wright2010,Cutri2014}
using a matching radius of $5\arcsec$, and identified 50 X-rays sources with 50 WISE objects. 
We supplement these with the remaining WISE point sources within the \xmm\ field of view  (Fig \ref{ircmd}).  
The WISE mid-IR color-color and color-magnitude diagrams show sources with distinct IR excesses: $W2-W3 > 1.1$~mag and $W1-W2 > 0.3$~mag.  
We  selected the sources with $W2 \le 12$~mag as YSOs; fainter sources are likely to be extragalactic 
\citep{Koenig2012,Koenig2014}. 
Two objects with $W1-W2 > 1.5$~mag are identified as protostars; the remainder as stars with disks 
(Class II; \citealp{Koenig2012}, \citealp{Fischer2016}).  We further identify sources with  $W2-W3 > 1.5$~mag, $W2 \le 12$~mag, 
and $W1-W2 < 0.3$~mag as transition disks.  X-ray sources with $W2-W3 < 1.1$~mag and $W2 \le 12$~mag were classified 
as disk-less pre-main sequence stars (Class III objects).
Another 12 objects with X-ray emission have near-IR colors and magnitudes similar to young pre-main sequence stars, i.e., 
no detections in $W3$ and $W4$ bands, and $W1-W2$ colors less than 0.3~mag (Fig \ref{ircmd}); these are classified as Class III objects.
Finally, we required $J-Ks > 0.47$ for sources detected in these bands; 
this eliminated one Class III source with a color bluer than that expected for pre-main stars (Fig. \ref{ircmd}).
The  {\em All}WISE images were visually inspected  as for the quality flags, signal to noise estimates, and $\chi^2$ values of the selected YSOs.  
Two X-ray sources that satisfy the criteria for transition disks are affected by artifacts from V1818 Ori; 
these are reclassified as  Class III objects. 
Four additional X-ray sources are found with $W2$ between 13.6 and 14.6~mag, $W2-W3 < 2$~mag (using  upper limits in the $W3$ band), 
and J-band magnitudes fainter than 14.8~mag.  We exclude these from our catalog for two reasons.  
First, these stars show $J$-band magnitudes below the Hydrogen burning limit for 3~Myr pre-main sequence stars at distance up to 400 pc 
\citep{Baraffe1998}.  
Second, the spatial position of such objects are more spread than the group of YSOs surrounding 
V1818 Ori, thus as a conservative choice we excluded them.  
Future spectroscopic observations are needed to determine whether they are truly pre-main sequence stars.  
The remaining X-ray sources with IR counterparts are much fainter $\ge 16$~mag and are not considered members.
}
About 41 X-ray sources are left without IR counterparts.  In \citet{Pillitteri2016} 
47/238 (in two XMM fields) were left without IR counterparts, i.e. about 23-24 per XMM field. 
Here we have almost twice the number of X-ray sources without IR matches. 
An inspection of the XMM image revealed that most of them have blue colors, this means their spectra are either hard
or heavily absorbed. They are scattered in the XMM field of view while the YSOs are concentrated 
toward the center. 
Taken together, the spatial distribution and spectral hardness indicate a background population.
The presence of a background cluster of galaxies is not excluded, the two extended sources 
found in the image could be part of it, although we cannot find a galaxy cluster listed in the 2MASS \citep{Tully2015} and 
SDSS catalogs \citep{Tempel2012}  within 20 arcmin from V1818 Ori.

Upper limits to rates were calculated at a threshold consistent with the value used for source detection.
The conversion factor (CF) between rates and unabsorbed fluxes (0.3-8.0 keV band) was derived with PIMMS ver. 4.9.
We assumed a thermal coronal spectrum described by an APEC model with $kT=1$ keV,  $Z=0.2 Z_\odot$. The average absorption
was estimated from an extinction map\footnote{\url{http://irsa.ipac.caltech.edu/applications/DUST/}} \citep{Schlafly2011},
and a value of $E(B-V)\sim1$ from which we derive $N_H=6.3\times10^{21}$ cm$^{-2}$ (corresponding to 
$A_V\sim3.3$ mag, for $R_V=3.1$, \citealp{Cardelli89},  value reported in the reddening vector in Fig. \ref{ircmd}). 
With these choices the CF has a value of $2.078\times10^{-11}$ erg cm$^{-2}$. 
Local higher extinction
and a different plasma temperature can change the CF by about 50\%. 
Given the low count statistics of the sources and the short exposure time it is impossible to perform a more accurate 
spectral analysis to derive their temperatures and fluxes. 

\section{Results and discussion  \label{results}}
{ We identified 31 members in the V1818 Group:  2 protostars (one with X-ray detection), 15 Class II sources 
(11 with X-ray detections), one X-ray undetected transition disk object, and 13 X-ray identified Class III sources 
(Table  \ref{ysoslist}).
Due to spatial overlap with a nearby source in the {\em ALL}WISE data, one of the Class II sources without an X-ray detection 
is considered tentative { (see Table 1 notes)}. The disk fraction among the YSOs with a X-ray detection is $0.46\pm0.14$.
\citet{Chiang2015} identified 24 YSOs in the V1818 Ori group using a combination of WISE colors and near-IR observations 
with the Subaru telescope. Twelve out of 24 YSOs selected by Chiang et al. are in common with our selection of members of the V1818 Ori group. 
The remaining 12 YSOs are not recovered by our criteria since they have WISE magnitudes below our limits or were 
identified in higher angular resolution, deeper near-IR observations.
Table \ref{ysoslist} lists the {\em 31 YSOs} identified as members of the V1818 Ori group.
 
These 31 YSOs including V1818 Ori form a group of pre main sequence stars 
coincident with the eastern edge of the dust ring surrounding \kori. 
}

{  From the fluxes listed in Table \ref{ysoslist}, we calculated luminosities and
we used a Kaplan-Meier estimator (as implemented in the R package {\em survival} in the case of left-censored data) 
to assess the distribution of the X-ray luminosities (XLFs) of the V1818 Ori's group on a set 
of distances between 115 pc and 1000 pc and 
used the Kolmogorov-Smirnov (KS) test to compare with the XLF of the YSOs found on the western edge of the \kori' ring. 
This is a relative distance analysis, but having assessed the distance to the Kappa Ori's YSOs in \citet{Pillitteri2016} this analysis
will be also calibrated in an absolute scale. 
 The Kappa Ori' XLF extends up to $10^{31}$ \lxu and is more skewed than the one of V1818 Ori.
Since the number of points in the two XLFs differ by a factor of $\sim8$, we can speculate that the high
luminosity tail in Kappa Ori' XLF could be due to flaring sources or to few foreground objects contaminating the sample.

The KS statistic peaks at a distance of $\sim270$ pc and the 90\% confidence range is $\sim230-350$ pc.
For  d$=400$ pc the KS statistic is $\sim0.06$ and for 900 pc the KS statistic is $\sim2\times10^{-6}$.
A distance of $d=900$ pc appears unrealistic, thus we can safely exclude that V1818 Ori and its surrounding 
YSOs belong to the Mon R2 association { as suggested by \citet{Chiang2015}.}
A distance of $d=400$ pc still appears higher than expected as the two XLFs overlap very marginally at 
the high luminosity tails, while $d\sim270$ pc seems the most reasonable.}
V1818 Ori and the surrounding YSOs seem unrelated to Orion A and L1641 and to the Mon R2 association. 
Rather, they are likely part of the \kori\ ring as they form a clump inside the eastern edge of it, 
analogous to the groups of YSOs identified on the 
western edge of the ring \citep{Pillitteri2016}. 

Based on the low background exposure time of 25 ks, 
the completeness limit is estimated above $L_X>10^{29}$ \lxu at a distance of 270 pc. 
This could leave undetected some very low mass stars. 
Assuming a saturated X-ray luminosity and a ratio $L_{X}/L_{bol}\sim 10^{-3}$, we have detected 
stars with $L_{bol}\ge 10^{32}$ \lxu or about $0.1 L_{bol,\odot}$, this value corresponds to spectral types
of M5, M4 and M2, respectively, when adopting isochrones at 1, 5 and 10 Myr \citep{Siess2000}.
The fraction of undetected YSOs with IR excesses is about 33\%, 
this alone suggests that the average distance of these objects is unlikely to be 900 pc as 
for the Mon R2 region.

{ The IR classification revealed only one protostar or Class I object in the \xmm\ FOV.
The paucity of protostars and the fraction of stars with disks suggest an age of  $2-5$ Myr, 
similar to that estimated for the YSOs on the western edge of the ring of \kori\ 
and slightly younger than the age of \kori\ itself  \citep{Pillitteri2016}.  
}

Considering the whole population of stars within $2\deg$ of \kori, its most massive member is \kori\ 
itself (a B0 star), then V1818 Ori is the second massive known member, followed by several low mass, M type stars.
Initial results from an optical spectroscopic survey of the YSOs identified in 
\citet{Pillitteri2016} reveal that these are predominantly low mass, M-type objects (Pillitteri et al., in prep.). 
We speculate that the group of YSOs around V1818 Ori is composed mostly of M type stars as well.
Including the sources from \citet{Pillitteri2016}, about 152 YSOs formed in the past $2-5$ Myrs within
the \kori\ ring, resulting in a rate of star formation of 30-75 stars Myr$^{-1}$. 

\begin{figure}
\centering 
\resizebox{0.49\textwidth}{!}{\includegraphics{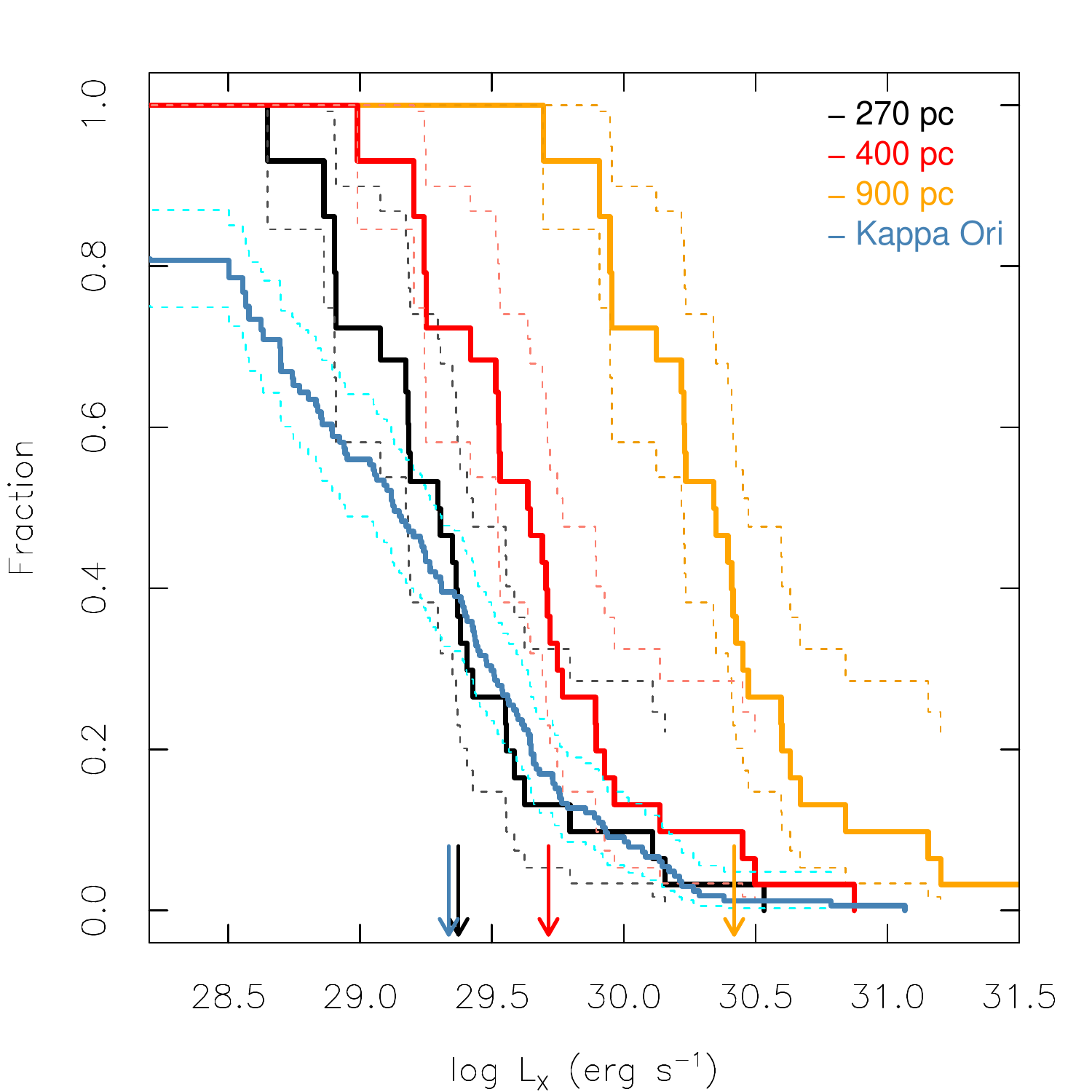}}
\caption{\label{xlf} Kaplan-Meier estimators of the X-ray Luminosity Function of the YSOs in V1818 Ori calculated for 
three distances: 900 pc (orange), 400 pc (red) and 270 pc (black). The curves at the 95\% confidence range are 
plotted in dashed lines. {  The XLF of the YSOs associated with Kappa Ori  \citep{Pillitteri2016} is shown for comparison.}
The arrows mark the medians of the different XLFs. The best agreement to the XLF of Kappa Ori group (distance 250--280 pc) 
is obtained for a distance of 270 pc.}
\end{figure}

\begin{table*}[t!]
\caption{List of identified YSOs and their 2MASS, WISE and X-ray photometry.  
Objects in common with \citet{Chiang2015} are indicated. Detection significance is given in units of the local 
background $\sigma$. Unabsorbed fluxes are given in the 0.3-8.0 keV band.  
For undetected YSOs the rates and fluxes are upper limits. Notes indicate  
contamination from spikes or bright nearby objects and names of objects known in literature.} 
\resizebox{1.00\textwidth}{!}{
\centering
\begin{tabular}{rrrrrrrrrrrrrrrrrl}
  \hline
  \hline
   N.& RA & Dec & designation & J & H & Ks & W1 & W2 & W3 & W4 & Classification & Chiang Id & Rate & Error & Signif & $\log$ Flux &  Note \\ 
     & J2000 (deg) & J2000 (deg) & & \multicolumn{7}{c}{mag}   &                &           & \multicolumn{2}{c}{ct ks$^{-1}$} & $\sigma_\mathrm{bkg}$ & & \\ 
  \hline
   1 & 88.29435 & -10.5443 & J055310.64-103239.6 & 13.9 & 13.0 & 12.8 & 12.5 & 12.4 & 12.1 & 8.08 &   Class III &   - & 7.09 & 0.55 & 20.60 & -12.83 &   \\ 
   2 & 88.32193 & -10.5394 & J055317.26-103221.8 & 13.4 & 12.5 & 12.1 & 11.6 & 11.2 &  9.5 & 7.67 &   Class II  &   - & 1.27 & 0.24 & 9.10  & -13.58 &  \\ 
   3 & 88.29679 & -10.5301 & J055311.22-103148.2 & 13.8 & 12.8 & 12.6 & 12.4 & 12.2 & 11.2 & 7.89 &   Class III &   - & 1.41 & 0.30 & 5.90  & -13.53 &  \\ 
   4 & 88.54125 & -10.4813 & J055409.89-102852.7 & 14.1 & 13.1 & 12.8 & 12.6 & 12.4 & 12.0 & 8.72 &   Class III &   - & 2.12 & 0.46 & 8.80  & -13.36 &  \\ 
   5 & 88.49402 & -10.4679 & J055358.56-102804.3 & 12.6 & 11.9 & 11.8 & 11.7 & 11.7 & 11.3 & 8.64 &   Class III &   - & 3.45 & 0.36 & 14.30 & -13.14 &  \\  
   6 & 88.37462 & -10.4431 & J055329.90-102635.1 & 14.7 & 13.8 & 13.5 & 13.3 & 13.3 & 12.3 & 8.93 &   Class III &   - & 0.85 & 0.14 & 8.10  & -13.75 &  \\  
   7 & 88.42907 & -10.4383 & J055342.97-102617.9 & 13.0 & 12.0 & 11.6 & 10.7 & 10.0 & 7.99 & 5.84 &   Class II  &   7 & 1.23 & 0.16 & 12.70 & -13.59 &  \\  
   8 & 88.38821 & -10.4269 & J055333.17-102536.8 & 13.4 & 12.6 & 12.4 & 12.2 & 12.1 & 10.9 & 7.82 &   Class III &   - & 1.96 & 0.19 & 17.00 & -13.39 &  \\  
   9 & 88.42543 & -10.4237 & J055342.10-102525.3 & 12.6 & 11.9 & 11.7 & 11.6 & 11.5 & 9.03 & 7.59 &   Class III &   - & 0.66 & 0.12 & 7.90  & -13.86 &  W3 Contaminated \\  
  10 & 88.44868 & -10.4184 & J055347.68-102506.1 & 12.8 & 11.9 & 11.4 & 11.0 & 10.4 & 7.99 & 5.62 &   Class II  &  21 & 7.91 & 0.40 & 41.80 & -12.78 &  \\  
  11 & 88.44001 & -10.4141 & J055345.60-102450.7 & 12.9 & 11.7 & 10.8 &  9.4 & 8.53 & 6.13 & 3.62 &   Class II  &  20 & 0.44 & 0.11 & 5.50  & -14.04 &  \\  
  12 & 88.48618 & -10.4128 & J055356.68-102446.1 & 13.4 & 12.6 & 12.4 & 12.3 & 12.2 & 12.1 & 8.99 &   Class III &  -  & 0.82 & 0.17 & 6.80  & -13.77 &  \\  
  13 & 88.37511 & -10.4092 & J055330.02-102432.9 & 10.2 & 9.91 & 9.72 & 8.97 & 8.49 & 4.94 & 2.84 &   Class II  &   2 & 0.40 & 0.10 & 4.70  & -14.08 &  IRAS05510-1025  \\ 
  14 & 88.42729 & -10.4002 & J055342.54-102400.5 &  9.8 & 7.63 & 5.96 & 4.09 & 2.91 & 0.69 &-1.07 &   Class II  &   1 & 0.45 & 0.10 & 7.40  & -14.03 &  V1818 Ori \\ 
  15 & 88.41904 & -10.4009 & J055340.57-102403.0 & 12.6 & 11.1 & 10.1 & 8.69 & 7.67 & 4.81 & 1.52 &   Class II  &  15 & 1.47 & 0.20 & 10.50 & -13.51 &   \\ 
  16 & 88.41640 & -10.3975 & J055339.93-102350.9 & 13.1 & 11.6 & 10.6 & 9.57 & 8.82 & 6.59 & 3.22 &   Class II  &  14 & 0.25 & 0.08 & 5.00  & -14.29 &   \\
  17 & 88.43755 & -10.3894 & J055345.01-102321.9 & 14.1 & 12.7 & 11.5 & 10.5 & 9.66 & 7.21 & 4.39 &   Class II  &  19 & 0.85 & 0.14 & 9.60  & -13.76 &   \\  
  18 & 88.39081 & -10.3798 & J055333.79-102247.3 & 13.4 & 12.5 & 12.0 & 11.1 & 10.5 & 8.78 & 7.01 &   Class II  &   9 & 1.32 & 0.16 & 13.50 & -13.56 &   \\  
  19 & 88.43200 & -10.3793 & J055343.68-102245.3 & 10.9 & 10.3 & 10.1 &   10 & 9.98 & 9.28 & 6.76 &   Class III &   - & 18.80& 0.57 & 72.60 & -12.41 &  W3 Contaminated \\ 
  20 & 88.42676 & -10.3708 & J055342.42-102214.7 & 13.2 & 12.3 & 11.9 & 11.3 & 10.9 & 8.84 & 6.79 &   Class II  &   - & 2.32 & 0.27 & 19.20 & -13.32 &    \\  
  21 & 88.40408 & -10.3455 & J055336.97-102043.7 & 13.9 & 13.1 & 12.9 & 12.8 & 12.7 & 11.7 & 9.05 &   Class III &   - & 1.09 & 0.18 &  8.60 & -13.65 &  \\  
  22 & 88.50644 & -10.3202 & J055401.54-101912.7 & 14.3 & 13.4 & 13.1 & 12.6 & 12.5 & 12.0 & 8.53 &   Class III &   - & 1.98 & 0.28 & 11.20 & -13.39 &  \\ 
  23 & 88.37238 & -10.3046 & J055329.37-101816.4 & 9.27 & 8.65 & 8.55 & 8.43 & 8.47 &  8.4 & 8.41 &   Class III &   - & 1.29 & 0.19 &  9.20 & -13.57 &  \\  
  24 & 88.34882 & -10.2910 & J055323.71-101727.4 & 16.8 & 15.8 & 14.0 & 11.6 & 9.74 & 7.78 & 5.57 &   Protostar &   - & 0.84 & 0.16 &  7.70 & -13.76 &  \\ 
  25 & 88.43061 & -10.1900 & J055343.34-101124.0 & 14.3 & 13.6 & 13.3 & 13.1 &   13 & 12.0 & 8.34 &   Class III &   - & 1.11 & 0.30 &  4.60 & -13.64 &  \\  
  26 & 88.34597 & -10.2923 & J055323.03-101732.2 & 17.3 & 16.3 & 14.3 & 12.4 & 9.97 & 7.32 & 3.68 &  Protostars &   - & $<0.84$ & --   & --    & $<-13.76$  & \\  
  27 & 88.38743 & -10.4534 & J055332.98-102712.2 & 15.7 & 13.6 & 12.4 & 11.2 & 10.4 & 8.35 & 5.59 &   Class II  &   7 & $<0.45$ & --   & --    & $<-14.03$  & \\ 
  28 & 88.29652 & -10.5440 & J055311.16-103238.3 & 10.7 & 10.3 & 10.1 & 9.89 &  9.7 & 8.43 &  6.6 & Trans. Disk &   - & $<7.09$ & --   & --    & $<-12.83$  & \\  
  29 & 88.35579 & -10.4543 & J055325.39-102715.6 & 13.9 & 12.4 & 11.7 & 11.1 & 10.4 & 8.51 & 6.12 &   Class II  &   5 & $<0.46$ & --   & --    & $<-14.02$  & Tentative \\ 
  30 & 88.44192 & -10.5811 & J055346.06-103452.1 & 13.7 & 12.7 & 12.1 & 11.2 & 10.6 & 8.26 & 6.19 &   Class II  &   - & $<1.04$ & --   & --    & $<-13.67$  & \\ 
  31 & 88.35442 & -10.4584 & J055325.06-102730.1 & 13.1 & 11.8 &   11 & 10.1 & 9.35 & 6.92 &  4.8 &   Class II  &   4 & $<0.48$ & --   & --    & $<-14.00$  & \\  
   \hline
\end{tabular}
}
\label{ysoslist}
\end{table*}

\section{Conclusions   \label{conclusions}}
We have presented the results from the analysis of an \xmm\ observation of the group of young stellar objects around 
the variable young star V1818 Ori, which is associated with the the Mon R2 region at 900 pc. 
However, given that the group is coincident with the eastern edge of the ring of warm dust surrounding \kori\ 
at $\sim250$ pc, we investigated their X-ray properties to infer their distances and ages analogously to 
what was done in \citet{Pillitteri2016}.

About 91 point-like X-ray sources and two faint extended sources were detected in 0.3-8.0 keV
band. About 50 X-ray sources have an infrared counterpart in WISE catalog. 
We used {\em All}WISE and 2MASS photometry to identify 31 objects as members of the group, 
including 25 X-ray sources, and classify them as protostars, Class II, III and transition disks.
We calculated the X-ray fluxes and luminosities (0.3-8.0 keV band) for the  25/31 YSOs detected in X-rays 
and upper limits  to fluxes and luminosities for the remaining 6/31. { By using a set of different distances 
for calculating the luminosities and comparing the resulting X-ray luminosity function (XLF) to that of 
the YSOs west of \kori, we concluded that 900 pc is unrealistic and the group of V1818 Ori is not related to Mon R2. 
A distance of 400 pc is still marginally consistent with the XLF of \kori' group, and thus the group does not 
appear related to Orion A and L1641.
For a distance of 270 pc (90\% confidence interval in $230-350$ pc) we obtain our best agreement with the XLF of 
the YSOs west of \kori\ and the  median X-ray luminosity of COUP YSOs.} 
We conclude that V1818 Ori and its surrounding young stars were born 
in the eastern edge of the \kori\ ring and are part of it. 

\begin{acknowledgements}
IP acknowledges support from INAF, ASI and the ARIEL consortium.
SJW was supported by NASA contract NAS8-03060 (Chandra X-ray Center)
This publication makes use of data products from the Wide-field Infrared Survey Explorer (WISE) 
and the Two Micron All Sky Survey (2MASS) retrieved through the NASA / IPAC Infrared Science Archive.
\end{acknowledgements}


\end{document}